\documentclass[conference, anonymous]{IEEEtran}
\IEEEoverridecommandlockouts
\usepackage{cite}
\usepackage{amsmath,amssymb,amsfonts}
\usepackage{algorithmic}
\usepackage{graphicx}
\usepackage{textcomp}
\usepackage{comment}
\usepackage{xcolor}
\usepackage{multirow}
\usepackage[table]{xcolor}
\definecolor{rowgray}{gray}{0.92}  

\def\BibTeX{{\rm B\kern-.05em{\sc i\kern-.025em b}\kern-.08em
    T\kern-.1667em\lower.7ex\hbox{E}\kern-.125emX}}
\begin{document}

\title{LLM-based Multi-class Attack Analysis and Mitigation Framework in IoT/IIoT Networks}

\author{\IEEEauthorblockN{Seif Ikbarieh}
\IEEEauthorblockA{\textit{Department of Computer Science} \\
\textit{Tennessee Tech University}\\
Cookeville, USA \\
smikbarieh42@tntech.edu}
\and
\IEEEauthorblockN{Maanak Gupta}
\IEEEauthorblockA{\textit{Department of Computer Science} \\
\textit{Tennessee Tech University}\\
Cookeville, USA \\
mgupta@tntech.edu}
\and
\IEEEauthorblockN{Elmahedi Mahalal}
\IEEEauthorblockA{\textit{Department of Electrical and Computer} \\
\textit{Engineering and Computer Science} \\
\textit{University of New Haven, USA}\\
emahalal@newhaven.edu}
}
\maketitle

\begin{abstract}
The Internet of Things has expanded rapidly, transforming communication and operations across industries but also increasing the attack surface and security breaches. Artificial Intelligence plays a key role in securing IoT, enabling attack detection, attack behavior analysis, and mitigation suggestion. Despite advancements, evaluations remain purely qualitative, and the lack of a standardized, objective benchmark for quantitatively measuring AI-based attack analysis and mitigation hinders consistent assessment of model effectiveness. In this work, we propose a hybrid framework combining Machine Learning (ML) for multi-class attack detection with Large Language Models (LLMs) for attack behavior analysis and mitigation suggestion. After benchmarking several ML and Deep Learning (DL) classifiers on the Edge-IIoTset and CICIoT2023 datasets, we applied structured role-play prompt engineering with Retrieval-Augmented Generation (RAG) to guide ChatGPT-o3 and DeepSeek-R1 in producing detailed, context-aware responses. We introduce novel evaluation metrics for quantitative assessment to guide us and an ensemble of judge LLMs, namely ChatGPT-4o, DeepSeek-V3, Mixtral 8x7B Instruct, Gemini 2.5 Flash, Meta Llama 4, TII Falcon H1 34B Instruct, xAI Grok 3, and Claude 4 Sonnet, to independently evaluate the responses. Results show that Random Forest has the best detection model, and ChatGPT-o3 outperformed DeepSeek-R1 in attack analysis and mitigation.
\end{abstract}


\section{Introduction}
The Internet of Things (IoT) is a large network of interconnected devices that communicate over the internet to deliver valuable data. The adoption of IoT has surged over the past decade, transforming operations across sectors. In 2024, the global IoT market was valued at about \$714.48 billion and is projected to reach \$4.06 trillion by 2032, growing at a 24.3\% Compound Annual Growth Rate (CAGR) \cite{fortune2025}. Within the realm of IoT, Industrial IoT (IIoT) integrates IoT into industrial processes, boosting efficiency in manufacturing, energy, healthcare, transportation, and many other industries.
The growth of IoT also expands the attack surface for cybercriminals. IoT devices often handle sensitive data, and their interconnected nature means a single vulnerability can compromise an entire network. Attacks on IoT have increased by 124\% in 2024 alone \cite{sonicwall2025}, underscoring the need for robust security measures to protect such environments, supporting the efficient management and secure growth of connected technologies. Clearly, securing these IoT-driven ecosystems is critical as our society embraces digital transformation, which is the core contribution of this work. 

Artificial Intelligence (AI) has become important in modern cybersecurity, enhancing defense mechanisms and reducing financial impact. Organizations using AI-based security prevention save an average of \$2.22 million over the organizations that did not deploy such technologies \cite{ibm2025}. Machine Learning (ML) detects attacks by analyzing manually selected features using algorithms such as Random Forest (RF) and Support Vector Machines (SVM). Deep Learning (DL) learns to extract hierarchical features from raw data using neural networks such as Convolutional Neural Network (CNN) for spatial analysis and Long Short-Term Memory (LSTM) for classifying sequential patterns. Both ML \cite{churcher2021experimental} and DL \cite{tacsci2024deep, gueriani2024adaptive} have been widely used for IoT attack detection.

Despite significant advances in AI for IoT security, there exists several critical \textbf{challenges and research gaps}. Firstly, current ML and DL models excel at attack detection but cannot analyze attack behaviors or provide actionable mitigation suggestions, limiting their operational practicality. Secondly, while existing cybersecurity applications using Large Language Models (LLMs) have shown potential for attack behavior analysis and mitigation suggestion \cite{10.1145/3663408.3663424, juttner2023chatids}, they have relied solely on qualitative evaluation. This highlights the need for a benchmark that introduces standardized, objective metrics to quantitatively assess LLM-generated analyses and mitigation suggestions, driving the development of more effective AI-driven IoT security solutions. Motivated by those gaps, our contributions in this work are as follows:
\begin{itemize}
    \item We propose a novel hybrid framework combining ML-based attack detection with LLM-based reasoning to analyze attack behavior and generate tailored mitigations for 13 different attack types in IoT/IIoT networks.
    \item We benchmark multiple ML and DL classifiers on the Edge-IIoTset \cite{9751703} and CICIoT2023 \cite{s23135941} datasets, integrating the top performer into our framework.
    \item We leverage structured role-play prompt engineering with Retrieval-Augmented Generation (RAG) to guide ChatGPT-o3 \cite{openai2025} and DeepSeek-R1 \cite{guo2025deepseek} in detailed attack analysis and tailored mitigation suggestion.
    \item We introduce a novel set of scoring metrics, eight judge LLMs and human expert use to independently assess and compare the LLM-generated attack behavior analyses and mitigation suggestions, enabling quantitative performance measurement and objective effectiveness comparison between ChatGPT-o3 and DeepSeek-R1.
\end{itemize}
The remaining sections of this paper are structured as follows. Section \ref{sec:rel} reviews related work on AI-based IoT attack detection, LLM prompting for attack analysis and mitigation, and the use of LLMs as judges. Section \ref{sec:datasets} describes the Edge-IIoTset and CICIoT2023 datasets. Section \ref{sec:frame} outlines our framework, including attack detection, RAG pipeline, and prompt engineering approach. Section \ref{sec:eval} presents our evaluation metrics and results. Finally, section \ref{sec:summary} concludes the paper and discusses future research directions.

\section{Related works}
\label{sec:rel}
ML and DL classifiers have shown promising results for attack detection on IoT networks. The authors in \cite{churcher2021experimental} benchmark seven different ML models for IoT attack detection on the Bot-IoT dataset, where RF achieved 99\% accuracy for binary classification, and K-Nearest Neighbor (KNN) achieved 99\% accuracy for multi-class classification. Taşcı in \cite{tacsci2024deep} proposes an optimized 1D CNN model for IoT attack and malware detection. The model achieves an accuracy of 98.36\% on the CICIoT2023 dataset, 99.90\% on CIC-MalMem-2022 dataset, and 99.99\% on CIC-IDS2017 dataset. The authors in \cite{gueriani2024adaptive} propose an intrusion detection system (IDS) for IIoT environments using a hybrid LSTM-CNN-Attention model. Their approach integrates LSTM for temporal dependency modeling, CNN for spatial feature extraction, and an attention mechanism to enhance feature selection. The model is evaluated on the Edge-IIoTset dataset, and achieved 99.04\% accuracy for multi-class classification, outperforming other DL models.

LLMs have recently been explored for security tasks that go beyond detection. The authors in \cite{10.1145/3663408.3663424} introduce ShieldGPT, which applies prompt engineering methods on GPT-4 to examine Distributed Denial of Service (DDoS) traffic and generate device-specific defense strategies. ShieldGPT's four components, namely attack detection, traffic representation, domain knowledge injection, and role representation, allow LLMs to examine network traffic and propose mitigations. Additionally, the authors in \cite{juttner2023chatids} introduce ChatIDS, which is designed for IoT environments such as smart homes, and aims to make IDS outputs more accessible to non-expert users. ChatIDS takes alerts generated by Suricata, Snort, and Zeek, enriches them with metadata, and passes them to ChatGPT to generate explanatory text and suggest mitigations.

The use of LLMs as judges is an emerging direction for automating evaluative tasks such as scoring, ranking, and verification. Authors in \cite{gu2024survey} review this line of work, outlining a structured evaluation pipeline with prompt-based querying, model selection, and post-processing. To enhance evaluation quality, they highlight techniques such as prompt engineering, majority voting, and fine-tuning on representative datasets.

While ML and DL classifiers demonstrate promising performance in IoT attack detection, they lack the ability to analyze attack behaviors and provide actionable mitigation suggestions, limiting their practicality in real-world cybersecurity operations. LLM-based approaches address this by using prompt engineering for behavior analysis and mitigation suggestion, but their reliance on qualitative analysis limits the ability to quantitatively benchmark effectiveness in a standardized and objective manner. To address those limitations, there is a need for a benchmark that enables quantitative evaluation of LLM responses across multiple attack scenarios while ensuring applicability to a wide range of cybersecurity use cases. With LLMs emerging as judges for consistent, scalable cybersecurity evaluation, our work addresses the mentioned gaps by introducing a hybrid detection and reasoning framework, along with a novel set of evaluation metrics to quantitatively measure LLM performance in IoT network attack scenarios.

\section{Datasets}
\label{sec:datasets}
The Edge-IIoTset dataset \cite{9751703} was collected from a seven-layer testbed of heterogeneous IoT/IIoT devices, sensors, and communication protocols to replicate real-world networks. It includes normal and malicious traffic across 14 attack types, grouped into five categories, namely DDoS, information gathering, Man-in-the-Middle (MITM), injection, and malware attacks. Each attack type represents distinct threats with different methods and impacts. From 1,176 extracted features, 61 were selected, covering network traffic behavior, system processes, protocol services, and security events. The dataset includes categorical, numeric, and textual data, supporting multi-modal ML, DL, and LLM-based processing. Table \ref{tab:edgeiiotset_dataset_features} lists a subset of representative features, their descriptions, and data types. The testbed features diverse devices such as Raspberry Pi edge nodes, industrial Modbus controllers, wireless routers, and Software-Defined Networking (SDN) controllers, offering a rich environment for attack detection and analysis.

\begin{table}[!t]
\caption{Subset of Features in the Edge-IIoTset Dataset}
\label{tab:edgeiiotset_dataset_features}
\begin{center}
\begin{tabular}{|l|p{3.35cm}|l|}
\hline
\textbf{Feature Name} & \textbf{Description} & \textbf{Data Type} \\
\hline
ip.src\_host & Source IP address of the packet & Categorical \\
\hline
tcp.flags & TCP flags indicating packet status & Numeric \\
\hline
http.request.method & Type of HTTP request (e.g., GET, POST) & Categorical \\
\hline
icmp.transmit\_timestamp & Timestamp of ICMP packet transmission & Numeric \\
\hline
dns.qry.name & Domain name requested in a DNS query & Categorical \\
\hline
mqtt.msg & Message payload in an MQTT communication & Categorical \\
\hline
\end{tabular}
\end{center}
\end{table}

The CICIoT2023 dataset \cite{s23135941} was captured from a large-scale smart home testbed with 105 IoT devices, including smart speakers, cameras, sensors, lighting systems, network hubs, and Raspberry Pi nodes acting as both benign and malicious nodes. The dataset contains both normal and attack traffic, capturing intra-IoT threats where compromised devices target other IoT systems. The dataset includes 33 attack classes that represent common, high-impact IoT threats, grouped into seven categories. The categories are DDoS, DoS, reconnaissance, web-based, brute force, spoofing, and Mirai attacks. The dataset includes 47 features extracted from packet flow windows, representing traffic behavior, protocol interactions, and header-level attributes. The dataset supports a range of feature types including categorical and numerical, making it suitable for ML and DL-based IDS development. Table \ref{tab:ciciot2023_dataset_features} presents a subset of representative features, their definitions, and data types. Example devices in the dataset include Google Nest Mini, Philips Hue Bridge, D-Link cameras, and Zigbee/Z-Wave-based smart devices, offering a rich testbed for cybersecurity use case experimentation.

\begin{table}[!t]
\caption{Subset of Features in the CICIoT2023 Dataset}
\label{tab:ciciot2023_dataset_features}
\begin{center}
\begin{tabular}{|l|p{4.5cm}|l|}
\hline
\textbf{Feature Name} & \textbf{Description} & \textbf{Data Type} \\
\hline
flow\_duration & Packet's flow duration & Numeric \\
\hline
Protocol Type & Indicates the type of protocol (e.g., TCP, UDP, ICMP, IP) & Categorical \\
\hline
Tot sum & Sum of packet lengths in the flow & Numeric \\
\hline
ack\_count & Count of packets with ACK flag set in the same flow & Numeric \\
\hline
DNS & Indicates if DNS is the application-layer protocol & Categorical \\
\hline
Tot size & The length of the packet & Numeric \\
\hline
\end{tabular}
\end{center}
\end{table}

We utilize the datasets for attack detection, LLM prompting, and benchmarking LLM-generated responses. To ensure fairness, we extract samples corresponding to the 13 attack types common to both datasets, categorized in Table \ref{tab:common_attacks}. The square brackets in Table \ref{tab:common_attacks} indicate cases where a single attack class in the Edge-IIoTset maps to multiple attack classes in the CICIoT2023.

\begin{table*}[!t]
\caption{Attack Types in the Framework}
\label{tab:common_attacks}
\begin{center}
\begin{tabular}{|l|p{12.2cm}|}
\hline
\textbf{Attack Category} & \textbf{Attack Types} \\
\hline
\textbf{DDoS} & Transmission Control Protocol Synchronization Flood (TCP SYN Flood) [TCP Flood, SYN Flood], User Datagram Protocol Flood (UDP Flood), Hypertext Transfer Protocol Flood (HTTP Flood), Internet Control Message Protocol Flood (ICMP Flood) \\
\hline
\textbf{Information Gathering/Reconnaissance} & Port Scanning, Vulnerability Scanning, Operating System (OS) Fingerprinting/OS Scanning \\
\hline
\textbf{Spoofing} & MITM [Address Resolution Protocol (ARP) Spoofing, Domain Name System (DNS) Spoofing] \\
\hline
\textbf{Injection} & Cross-Site Scripting (XSS), Structured Query Language (SQL) Injection, Uploading \\
\hline
\textbf{Malware} & Backdoor, Password Cracking/Dictionary Brute Force \\
\hline
\end{tabular}
\end{center}
\end{table*}

\section{LLM-based Framework}
\label{sec:frame}
\begin{figure*}[!t]
    \centering
    \includegraphics[width=0.9\linewidth]{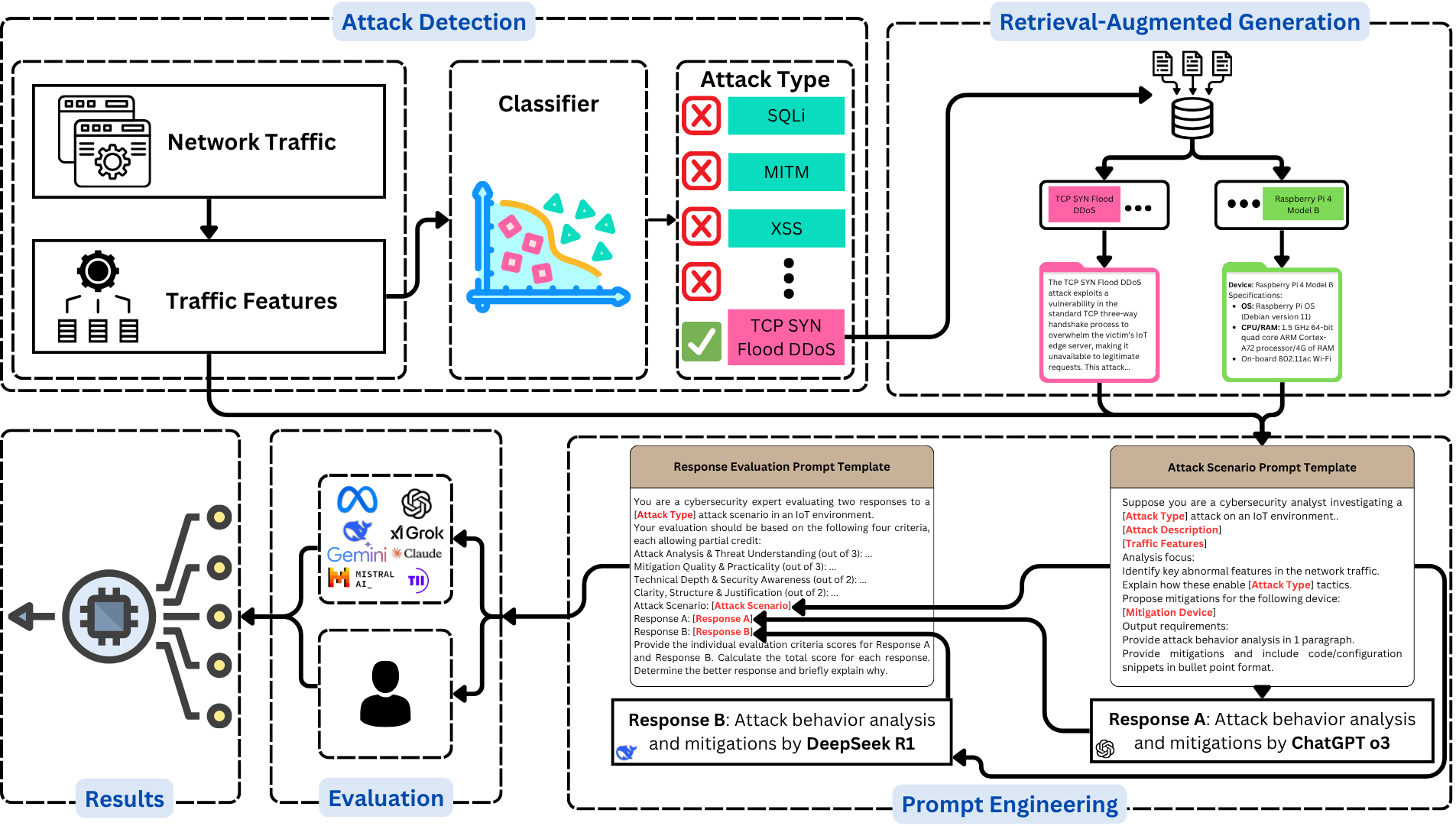}
    \caption{Overview of LLM-based Hybrid Framework}
    \label{fig:Framework-Diagram}
\end{figure*}

Our hybrid framework in Fig. \ref{fig:Framework-Diagram} integrates ML for attack detection with LLMs for behavior analysis and mitigation, comprising four components. The \textit{attack detection} component uses a trained classifier to identify the attack class from network traffic. The \textit{RAG} component retrieves attack descriptions and device information to enhance LLM prompts. In the \textit{prompt engineering} component, structured attack scenario prompts are sent to ChatGPT-o3 and DeepSeek-R1, which generate behavior analyses and mitigation suggestions. Those responses, along with the attack scenario, are embedded into the response evaluation prompt, which is designed for judge LLMs to objectively assess the responses. Finally, in the \textit{evaluation} component, the responses are evaluated by us and an ensemble of eight judge LLMs using our novel metrics to identify the most effective model. Our framework is implemented using Python, and all experiments were conducted on an Ubuntu 22.04 server running on an Intel(R) Xeon(R) Gold 5320 CPU @ 2.20GHz and an NVIDIA A100 40GB GPU.

\subsection{Attack Detection}
To classify attacks in IoT/IIoT networks, we benchmark 9 ML/DL models on the Edge-IIoTset and CICIoT2023 datasets to identify the most effective multi-class classifier for our framework. The ML models include RF, Extreme Gradient Boosting (XGB), KNN, SVM, Naïve Bayes (NB), and Logistic Regression (LR). DL models include CNN, LSTM, and Deep Neural Networks (DNN). LLMs are excluded from detection due to their high computational demands and the resource-constraints of IoT/IIoT environments.

We preprocess the data by removing irrelevant columns, handling missing values and duplicates, encoding categorical variables, and standardizing features, followed by an 80/20 train-test split. ML models use traditional supervised learning, while DL models leverage backpropagation and gradient optimization. This benchmarking compares feature-based and deep pattern learning, with the top performer selected as the detection model for our hybrid framework.


\subsection{Retrieval-Augmented Generation}
RAG enhances LLM outputs by retrieving relevant, context-specific information from an external knowledge base at inference time, and combining it with the LLMs' pre-trained knowledge to improve context awareness and reduce hallucinations \cite{fayyazi2024proverag}. Unlike static dictionaries, RAG supports flexible, effectively scalable retrieval as new attack types and input formats are introduced.

In our framework, RAG enriches LLM prompts with attack descriptions and device information. The attack knowledge base maps each label to a concise technical description from research papers and official sources such as the National Institute of Standards and Technology (NIST) \cite{8356377, doi:10.1177/1550147717741463, 9491117, app13105979, 10.5555/2206199}. These descriptions are extracted as a flat list of sentences and embedded for retrieval. The device knowledge base is similarly built using structured information from the datasets, including the Central Processing Unit (CPU) type, memory size, OS, and network interface, ensuring that the suggested mitigations are tailored to the constraints of the device.

We encode attack descriptions and device specifications with the all-MiniLM-L6-v2 sentence transformer, producing fixed-length vectors normalized for consistent similarity comparison and stored in Facebook AI Similarity Search (FAISS) indices for fast nearest-neighbor retrieval. At inference, the detected attack label and device name are encoded to query their respective FAISS indices, retrieving the most relevant description and specifications. This information is embedded into the LLM prompt, enabling responses that align with both the attack type and device specifications. While our RAG component receives attack labels, incorporating semantic similarity is essential to handle variations in attack class naming across datasets.

\subsection{Prompt Engineering}
After attack detection, LLMs are prompted to produce attack analyses and suggest mitigation strategies, which judge LLMs then assess using an evaluation prompt.

\subsubsection{Attack Behavior Analysis and Mitigation}
\begin{figure}[!t]
    \centering
    \includegraphics[width=1\linewidth]{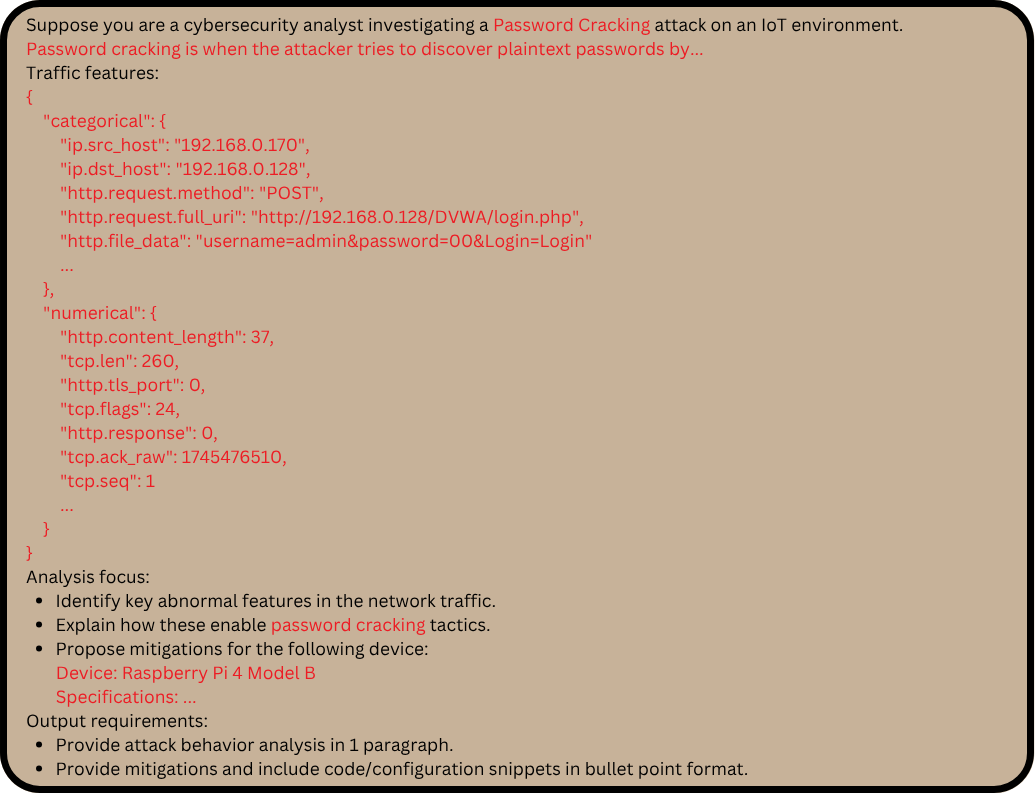}
    \caption{Edge-IIoTset Example Attack Scenario Prompt}
    \label{fig:Edge-IIoTset-Example-Attack-Scenario-Prompt}
\end{figure}

Our \textit{attack scenario prompt} is designed for attack behavior analysis and mitigation suggestion using role-play prompting and RAG. Using the ShieldGPT prompt template as a baseline, which only focused on DDoS attacks, we extend this approach to 13 attack types in our framework.
Fig. \ref{fig:Edge-IIoTset-Example-Attack-Scenario-Prompt} presents an example prompt for a Password Cracking attack, instructing ChatGPT-o3 and DeepSeek-R1 to analyze the attack behavior and suggest mitigations. We construct this attack scenario using a sample from the Edge-IIoTset. Reference \cite{guo2025deepseek} demonstrates the strong reasoning capabilities of ChatGPT-o1 and DeepSeek-R1, but we use ChatGPT-o3 as a direct successor to ChatGPT-o1 \cite{openai2025}, which has since been discontinued. We refer to ChatGPT-o3 and DeepSeek-R1 as the evaluated LLMs. The prompt assigns the evaluated LLMs the role of a cybersecurity analyst in an IoT environment, and explicitly states the attack class to maintain a focused perspective throughout the response. It includes JSON-formatted network traffic features from the attack instance, enabling LLMs to efficiently detect abnormal patterns, mimicking the way security professionals analyze traffic logs. The RAG component incorporates a brief description of the Password Cracking attack and the Raspberry Pi mitigation device specifications, allowing the LLMs to suggest mitigations tailored to the device’s constraints. The prompt concludes with structured output requirements, guiding the LLMs to generate organized attack analyses and mitigation suggestions to assist security professionals in understanding the attack and implementing countermeasures.

\begin{figure}[!t]
    \centering
    \includegraphics[width=1\linewidth]{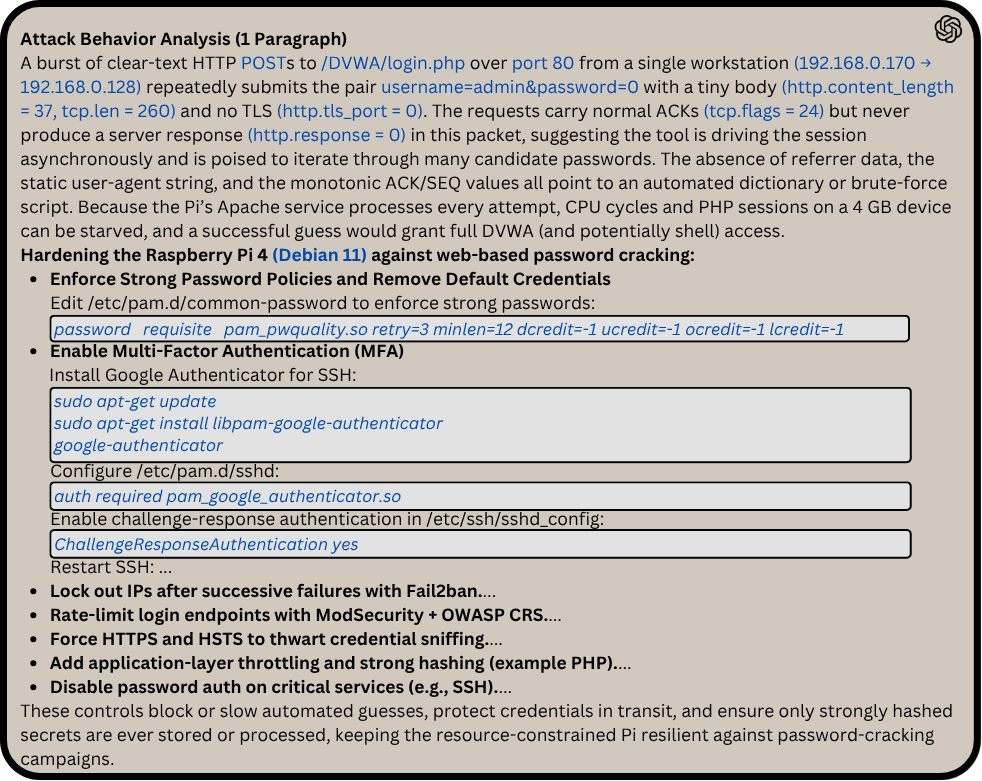}
    \caption{Edge-IIoTset Example ChatGPT-o3 Response}
    \label{fig:Edge-IIoTset-Example-ChatGPT-o3-Response}
\end{figure}

\begin{figure}[!t]
    \centering
    \includegraphics[width=1\linewidth]{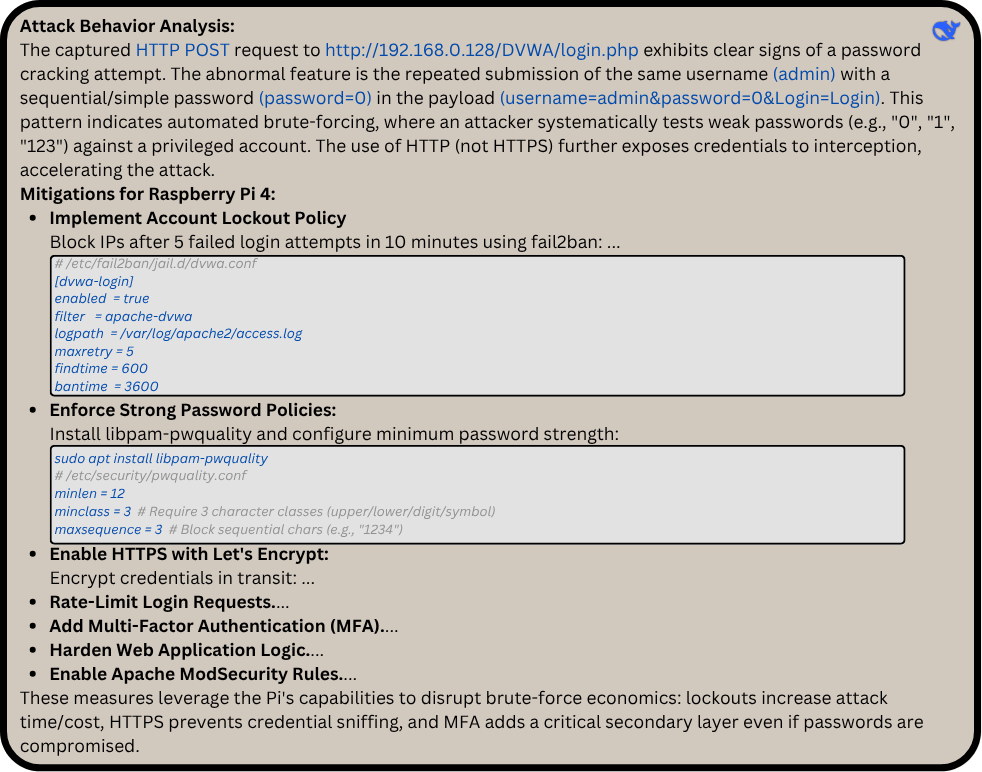}
    \caption{Edge-IIoTset Example DeepSeek-R1 Response}
    \label{fig:Edge-IIoTset-Example-DeepSeek-R1-Response}
\end{figure}

The attack scenario prompt is given to the evaluated LLMs to generate responses. Fig. \ref{fig:Edge-IIoTset-Example-ChatGPT-o3-Response} and \ref{fig:Edge-IIoTset-Example-DeepSeek-R1-Response} present the responses for the Password Cracking attack scenario from Fig. \ref{fig:Edge-IIoTset-Example-Attack-Scenario-Prompt}. Both responses include an attack behavior analysis that describes the adversary’s approach, highlighting key network indicators and attack patterns observed in the traffic data. The responses also include mitigation suggestions with code and configuration snippets, guiding security professionals in building countermeasures for the Password Cracking attack on the specified Raspberry Pi device. We used the same attack scenario prompt structure for the CICIoT2023, including the classified attack class, network traffic features in JSON format, and RAG-retrieved content. The evaluated LLMs’ responses exhibited a similar behavior as those for the Edge-IIoTset scenarios. however, due to space constraints, a full example for the CICIoT2023 is not shown. In this work, 28 versions of the attack scenario prompt were input into the evaluated LLMs, each corresponding to a specific attack class from the combined set of classes across both datasets, and the responses were recorded for evaluation.

\subsubsection{Response Evaluation with Judge LLMs}
\begin{figure}[!t]
    \centering
    \includegraphics[width=1\linewidth]{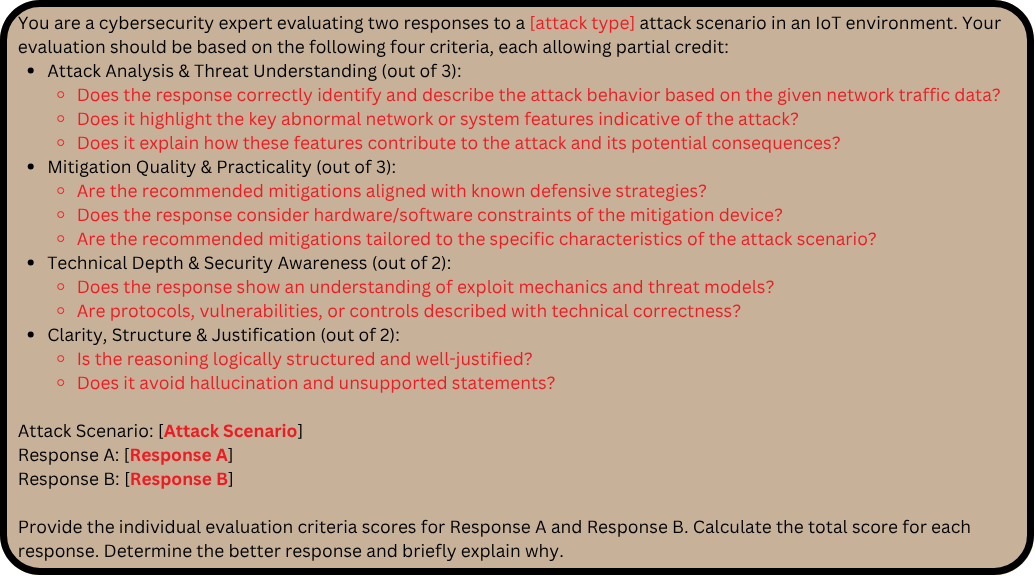}
    \caption{An Example Evaluation Prompt}
    \label{fig:Evaluation-prompt}
\end{figure}

Our \textit{response evaluation prompt}, shown in Fig. \ref{fig:Evaluation-prompt}, instructs the judge LLM to act as a cybersecurity professional scoring two responses for a given attack scenario. The evaluated LLMs' responses are displayed in an anonymous format to avoid biased judgment. The judge assesses each response on attack analysis and threat understanding, mitigation quality and practicality, technical depth and security awareness, as well as clarity, structure, and justification. Finally, the judge LLM scores each metric individually, provides the total score for each response, and briefly justifies which response is better.

We selected eight judge LLMs based on the CyberMetric benchmark and advancements of LLMs in cybersecurity knowledge \cite{10679494}. ChatGPT-4o, Mixtral 8x7B Instruct, Gemini 2.5 Flash, Meta Llama 4, and TII Falcon H1 34B Instruct were included as they are either featured in CyberMetric or successors to benchmarked models. Those models have been extensively benchmarked on various cybersecurity domains, including IoT network security. We also include newly released stable models, namely DeepSeek-V3, xAI Grok 3, and Claude 4 Sonnet, to enhance model diversity. To ensure unbiased evaluation, only one model per provider was chosen, covering varied model architectures. The evaluation prompt is used across all judge LLMs to assess the responses of the evaluated LLMs for all 28 attack class scenarios, allowing us to determine which evaluated LLM performed best in attack behavior analysis and mitigation suggestion.

\section{Performance Evaluation}
\label{sec:eval}
This section presents the metrics and results for identifying the best attack classifier and the most effective evaluated LLM for attack analysis and mitigation.

\subsection{Evaluation Metrics}
\label{sec:evaluation-metrics}
To quantify our framework’s effectiveness, we use two sets of evaluation metrics. The first set measures ML and DL classifier performance on the Edge-IIoTset and CICIoT2023 datasets using \textit{Precision}, \textit{Recall}, and \textit{F1-score}, allowing us to assess attack detection accuracy while minimizing misclassifications. The second set, utilized by us and the judge LLMs, evaluates LLM-generated responses through four metrics comprising 10 criteria, where each criterion is worth one point.

The first metric, \textit{Attack Analysis and Threat Understanding}, is scored on a 3-point scale and measures how accurately the evaluated LLMs identify and describe the attack from the given traffic data, detect key abnormal network or system features, and explain their role and potential impact. This is an important metric for determining which response provides an accurate and significant analysis of the attack scenario. The second metric, \textit{Mitigation Quality and Practicality}, is scored on a 3-point scale and assesses the relevance and applicability of suggested mitigations, their alignment with established cybersecurity practices, consideration of hardware and software limitations of the mitigation device, and suitability for the specific attack scenario. This metric is essential for assessing which response offers more practical and context-aware recommendations for real-world deployment in IoT/IIoT environments. The third metric, \textit{Technical Depth and Security Awareness}, is scored on a 2-point scale and assesses the evaluated LLMs’ understanding of exploitation mechanics, threat models, relevant protocols, vulnerabilities, and controls, ensuring technical accuracy and adherence to cybersecurity standards. The fourth metric, \textit{Clarity, Structure, and Justification}, is scored on a 2-point scale and measures whether the reasoning is well-structured, supported by sound logic, and free from hallucinations or unsupported claims. This metric reflects the importance of clear and justified analysis in effective cybersecurity communication.

This structured approach enables objective comparison of the generated attack analyses and mitigation strategy suggestions, providing a quantifiable basis for performance measurement to determine which evaluated LLM is better suited to enhance IoT/IIoT security within our hybrid framework.

\subsection{Results}
\subsubsection{Attack Classification Results}
Table \ref{tab:model_performance_comparison} presents the model classification performance results, where RF outperformed all models in multi-class attack detection across both datasets, achieving an F1-score of 0.9253 on the Edge-IIoTset and 0.8101 on the CICIoT2023. On the Edge-IIoTset, XGB followed closely with an F1-score of 0.9149, while KNN, SVM, and LR showed moderate results, and NB performed the worst. Among DL models, CNN and LSTM achieved moderate performance, whereas DNN struggled to capture the diversity of attack behaviors. On the CICIoT2023, XGB also ranked second with an F1-score of 0.6998, with the other ML and DL models showing limited effectiveness, reflecting challenges in handling noisy, variable IoT traffic data.


\begin{table}[!t]
\caption{Attack Classification Results}
\label{tab:model_performance_comparison}
\begin{center}
\begin{tabular}{|l|c|c|c|c|c|c|}
\hline
\multirow{2}{*}{\textbf{Model}} & \multicolumn{3}{c|}{\textbf{Edge-IIoTset}} & \multicolumn{3}{c|}{\textbf{CICIoT2023}} \\
\cline{2-7}
 & \textbf{P}$^{\mathrm{a}}$ & \textbf{R}$^{\mathrm{b}}$  & \textbf{F1}$^{\mathrm{c}}$  & \textbf{P}$^{\mathrm{a}}$  & \textbf{R}$^{\mathrm{b}}$  & \textbf{F1}$^{\mathrm{c}}$  \\
\hline
\rowcolor{rowgray}
RF       & 0.9320 & 0.9201 & \textbf{0.9253} & 0.9460 & 0.7514 & \textbf{0.8101} \\
\hline
XGB  & 0.9195 & 0.9128 & 0.9149 & 0.8277 & 0.6635 & 0.6998 \\
\rowcolor{rowgray}
\hline
KNN      & 0.7863 & 0.7594 & 0.7692 & 0.6548 & 0.5576 & 0.5821 \\
\hline
LR       & 0.8211 & 0.7332 & 0.7475 & 0.5629 & 0.4358 & 0.4576 \\
\rowcolor{rowgray}
\hline
LSTM     & 0.8644 & 0.7295 & 0.7292 & 0.6756 & 0.5622 & 0.5817 \\
\hline
SVM      & 0.7699 & 0.7169 & 0.7228 & 0.6977 & 0.5237 & 0.5444 \\
\rowcolor{rowgray}
\hline
CNN      & 0.7618 & 0.7149 & 0.7212 & 0.7131 & 0.5413 & 0.5607 \\
\hline
NB       & 0.7542 & 0.6824 & 0.6597 & 0.4882 & 0.4828 & 0.4079 \\
\rowcolor{rowgray}
\hline
DNN      & 0.6870 & 0.6569 & 0.6428 & 0.6582 & 0.5492 & 0.5658 \\
\hline
\multicolumn{5}{l}{$^{\mathrm{a}}$Precision, $^{\mathrm{b}}$Recall, $^{\mathrm{c}}$F1-score}
\end{tabular}
\end{center}
\end{table}

Tables \ref{tab:rf_classification_edgeiiotset} and \ref{tab:rf_classification_ciciot2023} present RF’s per-class performance on each dataset. RF achieved overall accuracies of 98.30\% on the Edge-IIoTset and 99.30\% on the CICIoT2023, with several perfect and near-perfect F1-scores for high-volume attacks such as flooding types. Performance was lower for subtle or infrequent attacks such as Uploading and XSS, where overlapping characteristics introduced ambiguity, causing misclassification. These results confirm RF as the optimal classifier for our framework, demonstrating strong adaptability to varying traffic characteristics and attack patterns in diverse IoT environments, while maintaining high accuracy, low false positive rates, and robust performance for subsequent LLM-based attack analysis and mitigation strategy suggestion.


\begin{table}[!t]
\caption{Classification Report (Edge-IIoTset dataset)}
\label{tab:rf_classification_edgeiiotset}
\begin{center}
\begin{tabular}{|l|c|c|c|c|}
\hline
\textbf{Attack Class} & \textbf{Precision} & \textbf{Recall} & \textbf{F1-score} & \textbf{Support} \\
\hline
\rowcolor{rowgray}
Backdoor & 0.9926 & 0.9808 & 0.9867 & 4,788 \\ \hline
HTTP Flood & 0.9299 & 0.9414 & 0.9357 & 9,684 \\ \hline
\rowcolor{rowgray}
ICMP Flood & 0.9999 & 0.9999 & 0.9999 & 13,717 \\ \hline
TCP SYN Flood & 1.0000 & 1.0000 & 1.0000 & 9,794 \\ \hline
\rowcolor{rowgray}
UDP Flood & 1.0000 & 1.0000 & 1.0000 & 24,274 \\ \hline
OS Fingerprinting & 0.8758 & 0.7016 & 0.7791 & 191 \\ \hline
\rowcolor{rowgray}
MITM & 1.0000 & 1.0000 & 1.0000 & 60 \\ \hline
Normal & 1.0000 & 1.0000 & 1.0000 & 272,927 \\ \hline
\rowcolor{rowgray}
Password Cracking & 0.8192 & 0.8358 & 0.8274 & 10,085 \\ \hline
Port Scanning & 0.9759 & 0.9995 & 0.9875 & 3,924 \\ \hline
\rowcolor{rowgray}
SQL Injection & 0.8298 & 0.8364 & 0.8331 & 10,327 \\ \hline
Uploading & 0.8278 & 0.7959 & 0.8115 & 7,393 \\ \hline
\rowcolor{rowgray}
Vulnerability Scanning & 0.9758 & 0.9615 & 0.9686 & 9,939 \\ \hline
XSS & 0.8209 & 0.8283 & 0.8246 & 2,894 \\ \hline
\rowcolor{rowgray}
\textbf{Macro Average} & 0.9320 & 0.9201 & 0.9253 & 379,997 \\ \hline
\textbf{Weighted Average} & 0.9830 & 0.9830 & 0.9830 & 379,997 \\
\hline
\rowcolor{rowgray}
\textbf{Accuracy (\%)} & \multicolumn{4}{c|}{\textbf{98.30}} \\
\hline
\end{tabular}
\end{center}
\end{table}

\begin{table}[!t]
\caption{Classification Report (CICIoT2023 Dataset)}
\label{tab:rf_classification_ciciot2023}
\begin{center}
\begin{tabular}{|p{2.6 cm}|c|c|c|c|}
\hline
\textbf{Attack Class} & \textbf{Precision} & \textbf{Recall} & \textbf{F1-score} & \textbf{Support} \\
\hline
\rowcolor{rowgray}
Backdoor & 0.9329 & 0.4203 & 0.5795 & 728 \\ \hline
Benign & 0.9154 & 0.9898 & 0.9511 & 243,322 \\ \hline
\rowcolor{rowgray}
HTTP Flood & 0.9981 & 0.9946 & 0.9964 & 6,450 \\ \hline
ICMP Flood & 1.0000 & 1.0000 & 1.0000 & 1,594,776 \\ \hline
\rowcolor{rowgray}
SYN Flood & 1.0000 & 0.9999 & 1.0000 & 900,820 \\ \hline
TCP Flood & 1.0000 & 1.0000 & 1.0000 & 996,211 \\ \hline
\rowcolor{rowgray}
UDP Flood & 1.0000 & 1.0000 & 1.0000 & 1,196,417 \\ \hline
DNS Spoofing & 0.8561 & 0.7994 & 0.8268 & 39,810 \\ \hline
\rowcolor{rowgray}
Dictionary Brute Force & 0.9773 & 0.5917 & 0.7371 & 2,983 \\ \hline
ARP Spoofing & 0.9381 & 0.8674 & 0.9014 & 67,988 \\ \hline
\rowcolor{rowgray}
OS Scanning & 0.8778 & 0.6446 & 0.7433 & 21,696 \\ \hline
Port Scanning & 0.8572 & 0.7030 & 0.7725 & 18,429 \\ \hline
\rowcolor{rowgray}
SQL Injection & 0.9563 & 0.4147 & 0.5785 & 1,160 \\ \hline
Uploading & 0.9636 & 0.2015 & 0.3333 & 263 \\ \hline
\rowcolor{rowgray}
Vulnerability Scanning & 0.9941 & 0.9970 & 0.9955 & 8,222 \\ \hline
XSS & 0.8685 & 0.3986 & 0.5464 & 878 \\ \hline
\rowcolor{rowgray}
\textbf{Macro Average} & 0.9460 & 0.7514 & 0.8101 & 5,100,153 \\ \hline
\textbf{Weighted Average} & 0.9929 & 0.9930 & 0.9926 & 5,100,153 \\ \hline
\rowcolor{rowgray}
\textbf{Accuracy (\%)} & \multicolumn{4}{c|}{\textbf{99.30}} \\
\hline
\end{tabular}
\end{center}
\end{table}

\subsubsection{Response Evaluation Results}
To illustrate our evaluation, we examine the responses in Fig. \ref{fig:Edge-IIoTset-Example-ChatGPT-o3-Response} and \ref{fig:Edge-IIoTset-Example-DeepSeek-R1-Response} for the Password Cracking attack scenario in Fig. \ref{fig:Edge-IIoTset-Example-Attack-Scenario-Prompt}. Both LLMs correctly identified the attack using HTTP traffic indicators such as request methods and headers. However, DeepSeek-R1 overlooked network-level features such as TCP flag patterns, limiting its explanation of how those features contribute to the attack’s mechanics and impact. As for the mitigation suggestions, both LLMs proposed widely known, functionally equivalent, and contextually appropriate defenses for the specified mitigation device, including Multi-Factor Authentication (MFA), strong password policy enforcement, and rate-limiting for account lockout \cite{mitreT1110002, owaspBruteForce}. In terms of technical depth, both LLMs demonstrated a solid understanding of the threat model, emphasizing its repetitive nature and automated brute force mechanics of the exploit. Additionally, the responses were clearly articulated, well-structured, and met the prompt’s formatting and justification requirements. Based on the evaluation criteria, ChatGPT-o3 scored a perfect 10 out of 10 while DeepSeek-R1 scored 8.5 out of 10 for this attack scenario.

\begin{figure}[!t]
    \centering
    \includegraphics[width=0.9\linewidth]{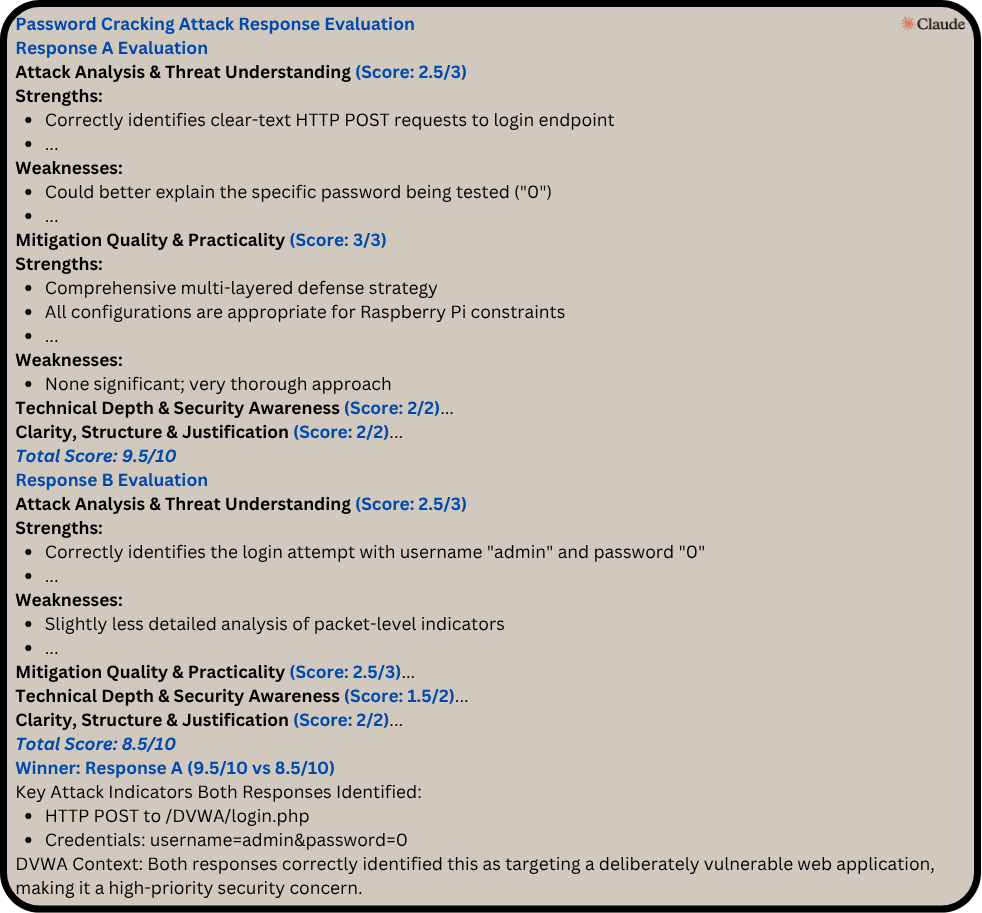}
    \caption{Judge LLM Response (Claude 4 Sonnet)}
    \label{fig:Claude-4-Sonnet-Response}
\end{figure}

Fig. \ref{fig:Claude-4-Sonnet-Response} shows the evaluation response from Claude 4 Sonnet, illustrating the judge LLM’s evaluative process, where it highlights gaps such as lack of specificity and insufficiently tailored mitigations in the evaluated LLMs’ responses. The evaluation reveals that Response A, generated by ChatGPT-o3, outperforms Response B, which corresponds to DeepSeek-R1, across all evaluation dimensions, receiving a total score of 9.5 out of 10 compared to 8 out of 10. The judge LLM provides a detailed justification for each score, noting that ChatGPT-o3 demonstrates deeper technical understanding of the exploit mechanics, highlights multiple attack vectors, and offers more robust, well-structured mitigations tailored to the Raspberry Pi 4 Model B environment. DeepSeek-R1, while accurate and well-organized, offers slightly less detailed analysis of low-level traffic indicators and provides a more surface-level interpretation of the attack's impact on system resources.

\begin{figure}[!t]
    \centering
    \includegraphics[width=0.9\linewidth]{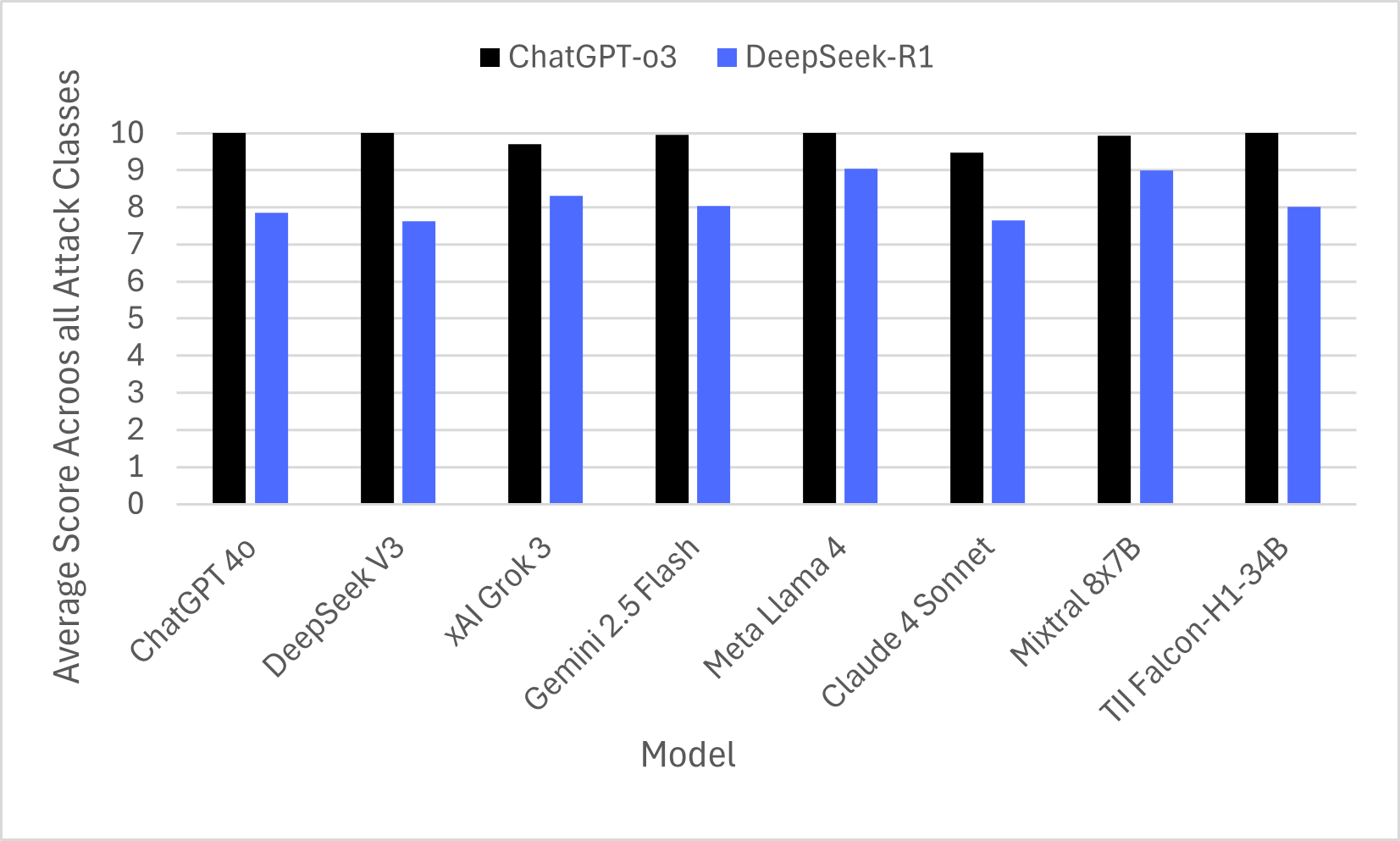}
    \caption{Evaluated LLM Performance (Edge-IIoTset Dataset)}
    \label{fig:edgeiiot_bar_chart}
\end{figure}

\begin{figure}[!t]
    \centering
    \includegraphics[width=0.9\linewidth]{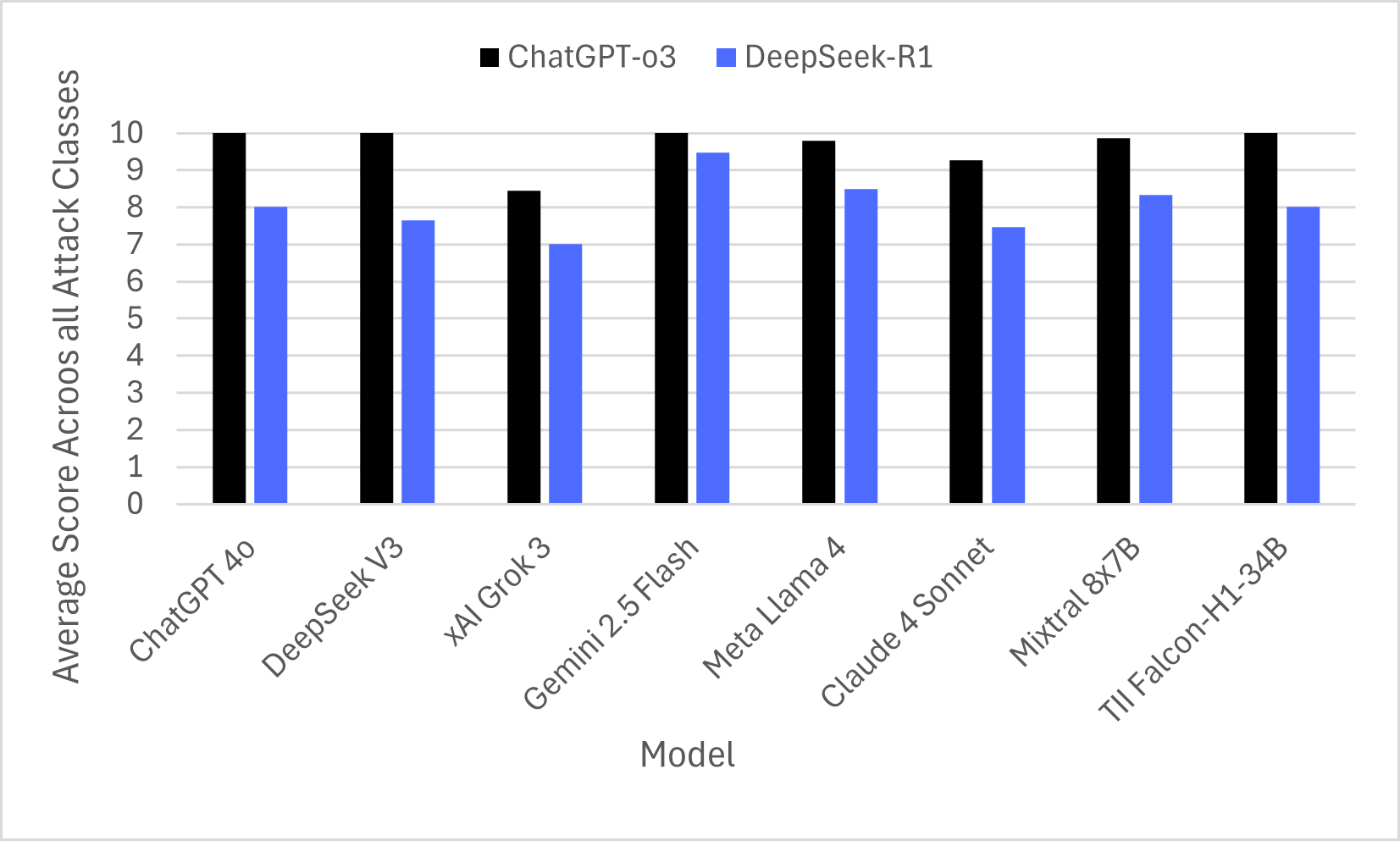}
    \caption{Evaluated LLM Performance (CICIoT2023 Dataset)}
    \label{fig:ciciot_bar_chart}
\end{figure}

We extended this evaluation process to every attack scenario from both datasets using the four metrics defined in Subsection \ref{sec:evaluation-metrics}, conducted jointly by us and the judge LLMs. Figures \ref{fig:edgeiiot_bar_chart} and \ref{fig:ciciot_bar_chart} present the average scores of the evaluated LLMs across all attack classes in the Edge-IIoTset and CICIoT2023 datasets, respectively, as scored by each judge LLM. On the Edge-IIoTset, ChatGPT-o3 consistently achieved higher scores, receiving perfect scores from ChatGPT-4o, DeepSeek-V3, Meta Llama 4, and TII Falcon-H1-34B-Instruct. In contrast, DeepSeek-R1’s scores exhibited greater variability, receiving lowest score of 7.62 from DeepSeek-V3. On the CICIoT2023, ChatGPT-o3 led again, earning perfect scores from ChatGPT-4o, DeepSeek-V3, Gemini 2.5 Flash, and TII Falcon-H1-34B-Instruct. The DeepSeek-R1 demonstrated lower performance, receiving the lowest score of 7 from xAI Grok 3. Given that DeepSeek-R1 scored lower with DeepSeek-V3 and showed nearly identical score differences between ChatGPT-4o and DeepSeek-V3 across both datasets, indicates no bias towards a peer model, especially with anonymized responses.

\begin{figure}[!t]
    \centering
    \includegraphics[width=0.9\linewidth]{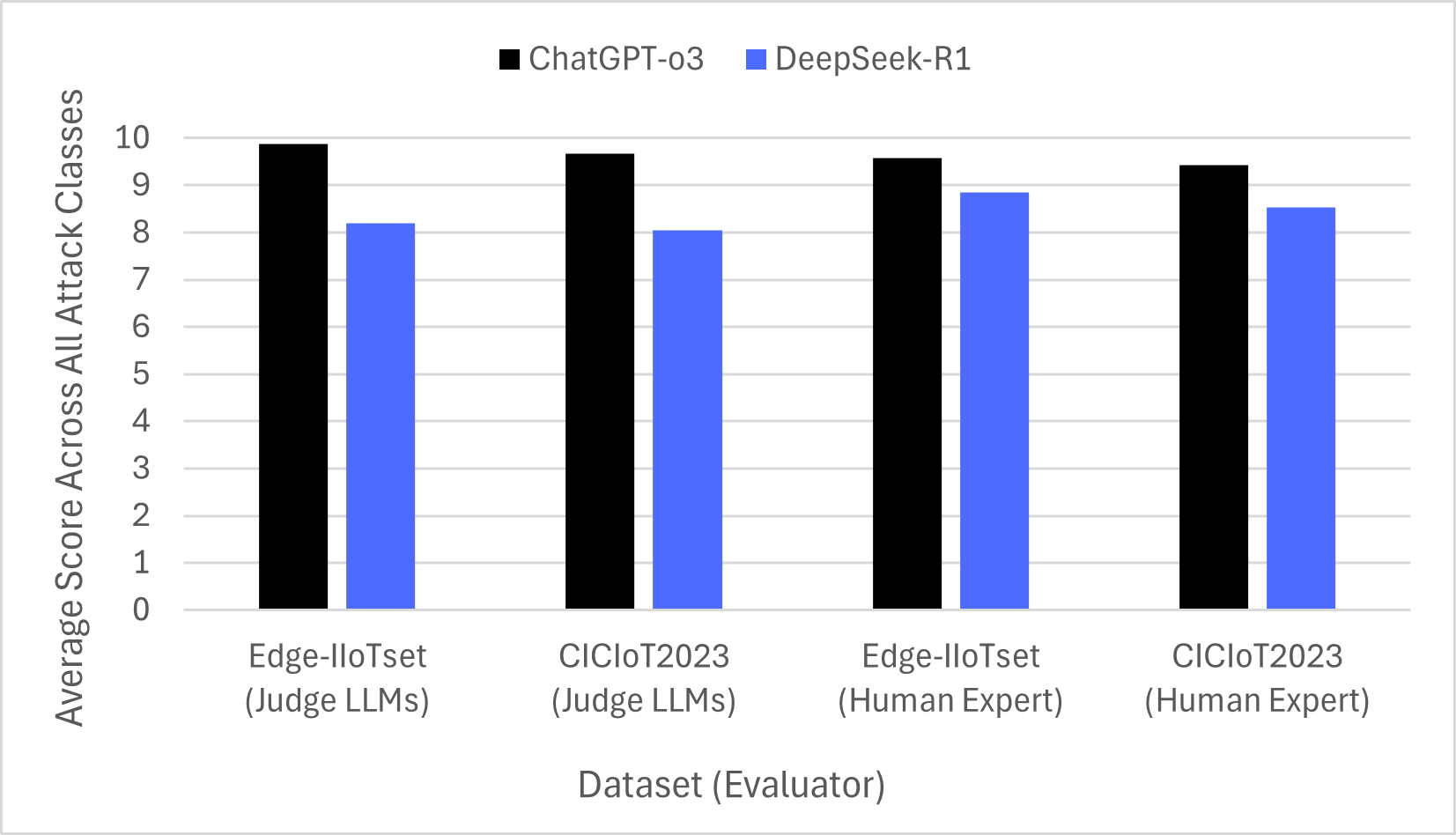}
    \caption{Overall Evaluated LLM Performance}
    \label{fig:overall-performance}
\end{figure}

Fig. \ref{fig:overall-performance} compares average scores for each evaluated LLM across all attack classes, providing a quantitative measurement of performance based on aggregated results from the ensemble of judge LLMs and human expert assessments. On the Edge-IIoTset, ChatGPT-o3 received an average score of 9.88 from judge LLMs and 9.58 from human expert, while DeepSeek-R1 scored 8.19 and 8.85, respectively. On the CICIoT2023, ChatGPT-o3 received an average score of 9.67 from judge LLMs and 9.43 from expert, while DeepSeek-R1 scored 8.05 and 8.53, respectively. These results show that ChatGPT-o3 delivers consistent, higher-quality attack analyses and mitigation suggestions across diverse IoT environments, while DeepSeek-R1’s lower scores reflect weaker traffic analysis depth and scenario adaptation. Overall, ChatGPT-o3 is the more effective reasoning model for our hybrid framework.


\section{Conclusion}
\label{sec:summary}
In this work, we presented a hybrid framework that integrates ML-based attack detection with LLM-driven attack behavior analysis and mitigation suggestion for IoT and IIoT networks. After identifying attacks using RF as the optimal multi-class classifier across both the Edge-IIoTset and CICIoT2023 datasets, we applied prompt engineering with RAG to guide ChatGPT-o3 and DeepSeek-R1 in generating detailed attack analyses and context-aware mitigation suggestions for the 13 common attack types in both datasets. To objectively evaluate the quality of the generated responses, we introduced a novel set of evaluation metrics and employed an ensemble of eight judge LLMs, along with human expert assessments. The results show that ChatGPT-o3 outperformed DeepSeek-R1 in providing technical, practical, and well-reasoned outputs across diverse attack types and IoT environments.

\section*{Acknowledgment}
Authors would like to thank Dr. Amr Hilal and Dr. Kshitiz Aryal for their insightful comments and suggestions which helped to improve the manuscript. This work is partially supported by the NSF grants 2416990 and 2346001, and NSA award H98230-24-1-0102 at Tennessee Tech University.

\bibliographystyle{ieeetr}
\bibliography{references}

\begin{thebibliography}{10}

\bibitem{fortune2025}
F.~B. Insights, ``Internet of things (iot) market report,'' 2025.
\newblock Accessed: Feb. 27, 2025.

\bibitem{sonicwall2025}
E.~Today, ``Smbs at critical risk, warns sonicwall as cyberattack speed
  surges,'' 2025.
\newblock Accessed: Feb. 27, 2025.

\bibitem{ibm2025}
I.~Security, ``Cost of a data breach report,'' 2025.
\newblock Accessed: Feb. 27, 2025.

\bibitem{churcher2021experimental}
A.~Churcher, R.~Ullah, J.~Ahmad, S.~Ur~Rehman, F.~Masood, M.~Gogate,
  F.~Alqahtani, B.~Nour, and W.~J. Buchanan, ``An experimental analysis of
  attack classification using machine learning in iot networks,'' {\em
  Sensors}, vol.~21, no.~2, p.~446, 2021.

\bibitem{tacsci2024deep}
B.~Ta{\c{s}}c{\i}, ``Deep-learning-based approach for iot attack and malware
  detection.,'' {\em Applied Sciences (2076-3417)}, vol.~14, no.~18, 2024.

\bibitem{gueriani2024adaptive}
A.~Gueriani, H.~Kheddar, and A.~C. Mazari, ``Adaptive cyber-attack detection in
  iiot using attention-based lstm-cnn models,'' in {\em IEEE International
  Conf. on Telecommunications and Intelligent Systems}, 2024.

\bibitem{10.1145/3663408.3663424}
T.~Wang, X.~Xie, L.~Zhang, C.~Wang, L.~Zhang, and Y.~Cui, ``Shieldgpt: An
  llm-based framework for ddos mitigation,'' in {\em Proceedings of the 8th
  Asia-Pacific Workshop on Networking}, APNet '24, (New York, NY, USA),
  p.~108–114, Association for Computing Machinery, 2024.

\bibitem{juttner2023chatids}
V.~J{\"u}ttner {\em et~al.}, ``Chatids: Explainable cybersecurity using
  generative ai,'' {\em arXiv preprint arXiv:2306.14504}, 2023.

\bibitem{9751703}
M.~A. Ferrag {\em et~al.}, ``{Edge-IIoTset: A New Comprehensive Realistic Cyber
  Security Dataset of IoT and IIoT Applications for Centralized and Federated
  Learning},'' {\em IEEE Access}, vol.~10, pp.~40281--40306, 2022.

\bibitem{s23135941}
E.~C.~P. Neto, S.~Dadkhah, R.~Ferreira, A.~Zohourian, R.~Lu, and A.~A.
  Ghorbani, ``Ciciot2023: A real-time dataset and benchmark for large-scale
  attacks in iot environment,'' {\em Sensors}, vol.~23, no.~13, 2023.

\bibitem{openai2025}
OpenAI, ``Introducing openai o3 and o4-mini,'' 2025.
\newblock Accessed: June 16, 2025.

\bibitem{guo2025deepseek}
D.~Guo, D.~Yang, H.~Zhang, J.~Song, R.~Zhang, R.~Xu, Q.~Zhu, S.~Ma, P.~Wang,
  {\em et~al.}, ``Deepseek-r1: Incentivizing reasoning capability in llms via
  reinforcement learning,'' {\em arXiv preprint arXiv:2501.12948}, 2025.

\bibitem{gu2024survey}
J.~Gu, X.~Jiang, Z.~Shi, H.~Tan, X.~Zhai, C.~Xu, W.~Li, Y.~Shen, S.~Ma, H.~Liu,
  {\em et~al.}, ``A survey on llm-as-a-judge,'' {\em arXiv preprint
  arXiv:2411.15594}, 2024.

\bibitem{fayyazi2024proverag}
R.~Fayyazi, S.~H. Trueba, M.~Zuzak, and S.~J. Yang, ``Proverag:
  Provenance-driven vulnerability analysis with automated retrieval-augmented
  llms,'' {\em arXiv preprint arXiv:2410.17406}, 2024.

\bibitem{8356377}
H.~Loi and A.~Olmsted, ``Low-cost detection of backdoor malware,'' in {\em 2017
  12th International Conference for Internet Technology and Secured
  Transactions (ICITST)}, pp.~197--198, 2017.

\bibitem{doi:10.1177/1550147717741463}
T.~Mahjabin, Y.~Xiao, G.~Sun, and W.~Jiang, ``A survey of distributed
  denial-of-service attack, prevention, and mitigation techniques,'' {\em
  International Journal of Distributed Sensor Networks}, vol.~13, no.~12,
  p.~1550147717741463, 2017.

\bibitem{9491117}
Z.~Marashdeh, K.~Suwais, and M.~Alia, ``A survey on sql injection attack:
  Detection and challenges,'' in {\em 2021 International Conference on
  Information Technology (ICIT)}, pp.~957--962, 2021.

\bibitem{app13105979}
I.~Alkhwaja, M.~Albugami, A.~Alkhwaja, M.~Alghamdi, H.~Abahussain, F.~Alfawaz,
  A.~Almurayh, and N.~Min-Allah, ``Password cracking with brute force algorithm
  and dictionary attack using parallel programming,'' {\em Applied Sciences},
  vol.~13, no.~10, 2023.

\bibitem{10.5555/2206199}
K.~A. Scarfone, M.~P. Souppaya, A.~Cody, and A.~D. Orebaugh, ``Sp 800-115.
  technical guide to information security testing and assessment,'' tech. rep.,
  National Institute of Standards \& Technology, 2008.

\bibitem{10679494}
N.~Tihanyi, M.~A. Ferrag, R.~Jain, T.~Bisztray, and M.~Debbah, ``Cybermetric: A
  benchmark dataset based on retrieval-augmented generation for evaluating llms
  in cybersecurity knowledge,'' in {\em 2024 IEEE International Conference on
  Cyber Security and Resilience}.

\bibitem{mitreT1110002}
{MITRE ATT\&CK}, ``Brute force: Password guessing — att\&ck t1110.002,''
  2025.
\newblock Accessed: June 19, 2025.

\bibitem{owaspBruteForce}
{OWASP Foundation}, ``Blocking brute force attacks - owasp,'' 2025.
\newblock Accessed: June 19, 2025.

\end{thebibliography}

\end{document}